\documentclass[conference]{IEEEtran}
\IEEEoverridecommandlockouts
\usepackage{cite}
\usepackage{xspace}
\usepackage{amsmath,amssymb,amsfonts}
\usepackage{newtxmath}
\usepackage{algorithmic}
\usepackage{graphicx}
\usepackage{textcomp}
\usepackage{enumitem}
\usepackage{hyperref}
\usepackage{cleveref}
\usepackage{lipsum}
\crefname{figure}{Fig.}{Figs.}
\crefname{table}{Table}{Tables}
\crefname{section}{Section}{Sections}
\Crefname{figure}{Figure}{Figures}
\usepackage[table]{xcolor}
\usepackage{booktabs}
\usepackage{siunitx}

\newcommand{\RQ}[1]{\textit{RQ}${}_{\mathrm{#1}}$}
\usepackage{framed}
\newcommand{\Conclusion}[1]{\begin{framed}\noindent #1\end{framed}}

\newcommand{\repository}[1]{\textit{#1}}
\newcommand{\refactoring}[1]{\textit{#1}}
\newcommand{\attribute}[1]{\textit{#1}}

\newcommand{\microchange}[1]{\textit{#1}}

\def\AddConditionalStatement{\microchange{Add Conditional Statement}\xspace}
\def\AddConjunctOrDisjunct{\microchange{Add Conjunct or Disjunct}\xspace}
\def\AdjustConditionBoundary{\microchange{Adjust Condition Boundary}\xspace}
\def\ConditionalToBooleanReturn{\microchange{Conditional to Boolean Return}\xspace}
\def\ConditionalToExpression{\microchange{Conditional to Expression}\xspace}
\def\ConditionalToSwitch{\microchange{Conditional to Switch}\xspace}
\def\ExtendElseWithIf{\microchange{Extend Else with If}\xspace}
\def\ExtendIfWithElse{\microchange{Extend If with Else}\xspace}
\def\FlipLogicOperator{\microchange{Flip Logic Operator}\xspace}
\def\MoveInwardCondition{\microchange{Move inward Condition}\xspace}
\def\MoveOutwardCondition{\microchange{Move outward Condition}\xspace}
\def\RemoveConditionalStatement{\microchange{Remove Conditional Statement}\xspace}
\def\RemoveConjunctOrDisjunct{\microchange{Remove Conjunct or Disjunct}\xspace}
\def\RemoveElse{\microchange{Remove Else}\xspace}
\def\ReverseCondition{\microchange{Reverse Condition}\xspace}
\def\SwapThenAndElse{\microchange{Swap Then and Else}\xspace}
\def\UnwrapStatementFromBlock{\microchange{Unwrap Statement from Block}\xspace}
\def\UnwrapStatementFromConditional{\microchange{Unwrap Statement from Conditional}\xspace}
\def\WrapStatementInBlock{\microchange{Wrap Statement in Block}\xspace}
\def\WrapStatementInConditional{\microchange{Wrap Statement in Conditional}\xspace}
        
\def\INS{\mathrm{INS}}
\def\DEL{\mathrm{DEL}}
\def\UPD{\mathrm{UPD}}
\def\MOV{\mathrm{MOV}}

\newcommand{\Circled}[1]{\textcircled{{\footnotesize #1}}}

\begin{document}

\title{Understanding Code Change with Micro-Changes}

\author{\IEEEauthorblockN{Lei Chen}
\IEEEauthorblockA{\textit{School of Computing}\\
\textit{Tokyo Institute of Technology}\\
Tokyo, Japan \\
chenlei@se.c.titech.ac.jp}
\and
\IEEEauthorblockN{Michele Lanza}
\IEEEauthorblockA{\textit{Software Institute}\\
\textit{Universit\`{a} della Svizzera italiana}\\
Lugano, Switzerland\\
michele.lanza@usi.ch}
\and
\IEEEauthorblockN{Shinpei Hayashi}
\IEEEauthorblockA{\textit{School of Computing}\\
\textit{Tokyo Institute of Technology}\\
Tokyo, Japan\\
hayashi@c.titech.ac.jp}
}

\maketitle

\pagestyle{plain}
\thispagestyle{plain}

\begin{abstract}
A crucial activity in software maintenance and evolution is the comprehension of the changes performed by developers, when they submit a pull request and/or perform a commit on the repository. Typically, code changes are represented in the form of code diffs, textual representations highlighting the differences between two file versions, depicting the added, removed, and changed lines.
This simplistic representation must be interpreted by developers, and mentally lifted to a higher abstraction level, that more closely resembles natural language descriptions, and eases the creation of a mental model of the changes. However, the textual diff-based representation is cumbersome, and the lifting requires considerable domain knowledge and programming skills.
We present an approach, based on the concept of {\em micro-change}, to overcome these difficulties, translating code diffs into a series of pre-defined change operations, which can be described in natural language. We present a catalog of micro-changes, together with an automated micro-change detector.
To evaluate our approach, we performed an empirical study on a large set of open-source repositories, focusing on a subset of our micro-change catalog, namely those related to changes affecting the conditional logic. We found that our detector is capable of explaining more than 67\% of the changes taking place in the systems under study.

\end{abstract}

\begin{IEEEkeywords}
software evolution, source code mining, diff analysis
\end{IEEEkeywords}

\section{Introduction}

Understanding the code changes made by developers during the submission of pull requests and/or commits to a repository is a critical activity in the maintenance and evolution of software~\cite{bacchelli2013expectations}. It can help developers track the state of the software system~\cite{sadowski2018modern}, maintain the current code~\cite{nguyen2009clone}, pinpoint when and where an issue is introduced~\cite{kim2009discovering}, and plan for future development. 

\begin{figure*}[t]\centering
    \includegraphics[width=\linewidth]{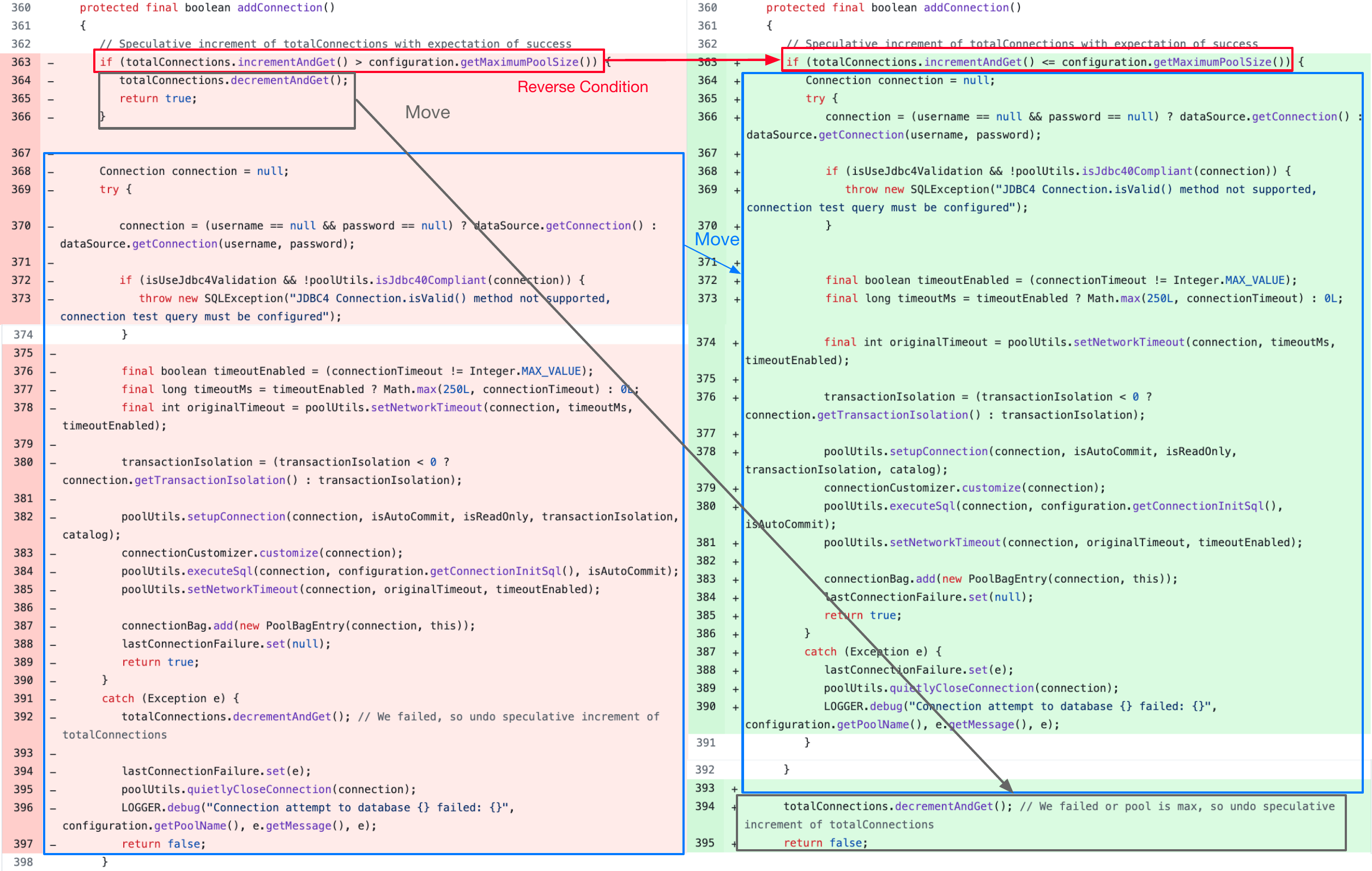}
    \caption{Example of the code diff in a commit in repository \repository{HikariCP}.}
    \label{f:example_micro_change}
\end{figure*}

The code changes are usually represented as textual code differences~(diffs), which provide a detailed account of which lines and characters of source code have been added, removed, or modified between two file versions. An example of code diff found in a commit in the open source repository~\repository{HikariCP}\footnote{\url{https://github.com/brettwooldridge/HikariCP/commit/2260cc2}} is shown in \cref{f:example_micro_change}. The side-by-side code comparison illustrates modifications, where the left side code represents the pre-update version and the right side displays the post-update version, with red highlighting deletions and green indicating additions. The diff view does little to help in understanding that the nearly 40 lines involved in this change were the result of the developer inverting a conditional expression, and subsequently moving a piece of code: there are too many trees depicted here to see the forest.

The diff view, while precise at a textual level, does not meet the cognitive needs of developers when understanding code changes~\cite{maletic2004supporting}, as developers do not think in terms of characters and lines, but at a higher abstraction level, forming a mental model of the logic behind the code~\cite{DBLP:journals/jss/StoreyFM99}. Developers need to compare the code on the left and right sides line by line and interpret the diffs by mentally lifting them to a higher abstraction level, closer to a natural language narrative that encapsulates the essence of the change operations~\cite{von1993code}. The lifting process is time and energy-consuming~\cite{DBLP:conf/sigsoft/SiegmundPPAHKBB17} and also requires extensive domain knowledge and proficient programming skills~\cite{DBLP:journals/tse/PeitekSAKPBLSB20}. Though commit messages are designed to reveal the code changes contained in commits, they often are of low quality and cannot convey the change well~\cite{tian2022makes}.

To overcome the above difficulty and mitigate the gap between the raw, textual-level diffs and the deeper, conceptual understanding developers need, we propose the concept of \textit{micro-changes}. Micro-changes are a set of code change operations described in natural language, designed to bridge the cognitive divide by translating the textual diffs into more understandable natural-language described operations. As opposed to code diffs, which operate at a low semantic level, focusing exclusively on the literal character alterations, micro-changes lift the semantic level of code diff by distilling their essence into comprehensible, natural-language descriptions. The contributions of this paper are:

\begin{itemize}
    \item The definition of micro-changes to model and explain code changes.
    \item A catalog of 20 types of conditional-related micro-changes.
    \item An automated micro-change detector.
    \item An empirical study on 73 open source Java repositories, finding that 67.1\% of the conditional-related code changes can be covered by the detected micro-changes.
\end{itemize}

\section{Related Work}

\subsection{Modern Code Review}

Code review is a key practice for ensuring software quality and facilitating maintenance, serving as checkpoint for identifying issues and fostering code improvements~\cite{ackerman1984software, ackerman1989software}. Modern code review is a lightweight, tool-based way focusing on code changes and is widely adopted for both industrial~\cite{sadowski2018modern, alomar2021refactoring, bacchelli2013expectations} and open source software~\cite{mcintosh2016empirical}.

This process typically starts with a developer submitting a pull request that involves single or multiple \textit{commits}, which is a set of changes, deletions or additions to the code encapsulated in a single update. These changes are presented as a \textit{diff} (difference), showing the alterations between the new and existing code, highlighting the additions, deletions, and modifications.

Other team members review the diff, providing feedback, suggestions, and approval, ensuring the code change is optimal in terms of quality, functionality, and adherence to project standards before it is merged into the main branch.

The diff being reviewed can be categorized into two types: \textit{text-based diff} and \textit{tree-based diff}.

\subsection{Text-Based Diff}

The text-based diff is widely used in computing differences between two versions of a source file~\cite{miller1985file,DBLP:journals/algorithmica/Meyers86}. It shows the added, deleted, and changed text lines.

Canfora et al.~\cite{canfora2007identifying,canfora2008tracking} introduced a line differencing methodology, termed \textit{ldiff}, that possesses the capability to monitor the positions of lines irrespective of the programming languages involved. This methodology initially employs the Unix diff algorithm to determine the lines that remain unchanged. Subsequently, it utilizes a combination of set-based and sequence-based metrics to facilitate the comprehensive mapping of the remaining lines. Asaduzzaman et al.~\cite{asaduzzaman2013lhdiff} proposed \textit{LHDiff}, which adopts the \textit{simhash}~\cite{charikar2002similarity} technique to speed up the mapping process and employs a set of heuristics to improve the effectiveness of tracking source locations.

However, the text-based diff approach fails to effectively leverage the inherent structure of source code.

\begin{table*}[ht]
    \centering
    \caption{Conditional-Related Micro-Change Catalog Overview}
    \label{t:catalog}
    \rowcolors{2}{gray!10}{white}
    \begin{tabular}{lp{13cm}}
        \toprule
        {\bf Micro-change} & {\bf Description}\\ \midrule
        \AddConditionalStatement & Add a new conditional statement. \\
        \AddConjunctOrDisjunct & Modify an existing conditional statement by appending an additional condition using logical AND (\texttt{\&\&}) or logical OR (\texttt{||}) operators to refine or expand the criteria for the statement's execution.\\
        \AdjustConditionBoundary & Modify a comparison operator in a conditional statement to alter its boundary condition.\\ 
        \ConditionalToBooleanReturn & Simplify a conditional statement that directly returns a boolean value by replacing the if statement with a direct return of the condition's evaluation.\\ 
        \ConditionalToExpression & Simplify an if-else statement into a concise conditional operator.\\
        \ConditionalToSwitch & Transform a series of if statements into a switch statement.\\
        \ExtendElseWithIf & Replace a simple else clause in an if-else statement with an else if condition, adding a new conditional check to the existing control flow for more specific action selection based on multiple conditions.\\
        \ExtendIfWithElse & Add an else clause to an existing if statement. \\
        \FlipLogicOperator & Alter the logical operator within a conditional statement, switching between AND ($\&\&$) and OR ($||$). \\ 
        \MoveInwardCondition & Reorganize nested conditional statements by moving one of the conditions from an outer if statement to be combined with the condition of an inner if statement. \\
        \MoveOutwardCondition & Reorganize nested conditional statements by moving one of the conditions from an inner if statement to be combined with the condition of an outer if statement. \\
        \RemoveConditionalStatement & Remove an existing conditional statement. \\
        \RemoveConjunctOrDisjunct & Modify an existing conditional statement by removing conditions from a compound logical expression that concatenates multiple conditions by logical AND (\texttt{\&\&}) or logical OR (\texttt{||}). \\
        \RemoveElse & Remove the else clause from an if-else statement, leaving only the if statement and its associated action.\\
        \ReverseCondition & Invert the logic of a conditional expression within an if statement, changing the condition to its logical opposite.\\
        \SwapThenAndElse & Swap the actions of the then and else clauses of an if-else statement.\\
        \UnwrapStatementFromBlock & Remove the curly braces from a block containing a single statement following an if statement.\\
        \UnwrapStatementFromConditional & Remove the conditional check around a statement, so the statement executes unconditionally, independent of the previously specified condition. \\       
        \WrapStatementInBlock & Enclose a single statement within curly braces following an if statement.\\ 
        \WrapStatementInConditional & Wrap an existing statement, within an if statement.\\
        \bottomrule
    \end{tabular}
\end{table*}

\subsection{Tree-Based Diff}

The tree-based diff utilizes the abstract syntax tree (AST) of the source code and tree differencing algorithms to extract more detailed change information.

Fluri et al.~\cite{fluri2007change} proposed a tree difference algorithm, which can find the matching nodes between two ASTs and calculate the minimum edit script that transforms one AST to another.

Faller et al.~\cite{falleri2014fine} proposed a tool named \textit{GumTree} that also takes the ``move'' action into account in addition to the insertion, deletion, and update actions. They evaluated their tool on a large-scale dataset to measure its accuracy.

However, the tree-based diff, while precise and detailed, requires knowledge about ASTs to understand.
Additionally, it is at a low abstraction level which is not intuitive and does not necessarily align with how developers conceptualize code changes.

\subsection{Refactoring}

Refactoring is the concept at a higher level of abstraction by emphasizing systematic modifications that enhance the internal structure of the code while preserving its external behavior~\cite{fowler2018refactoring,opdyke1992refactoring}, and it is a key practice in agile development processes~\cite{chen2016perspectives}.

To detect refactorings, Silva et al.\ proposed RefDiff~\cite{silva2017refdiff,silva2020refdiff}, which detects refactorings through two phases: source code analysis and relationship analysis. Initially, it parses the source code to construct a model of high-level entities, e.g., types, methods, and fields. Then, it assesses relationships within this model across code changes to identify refactorings by matching these relationships against predefined rules.

Tsantails et al.\ introduced RefactoringMiner~\cite{tsantalis2018accurate, Tsantalis:TSE:2020:RefactoringMiner2.0}, marking the first refactoring mining tool that operates without the need for any code similarity thresholds. This tool detects refactorings through AST-based statement matching with predefined rules. The code entities in the two revisions are matched in top-down order based on text similarity.

This concept aligns with the need for a more abstract representation of code changes. However, the inherent limitation of refactoring, its focus on behavior-preserving transformations, restricts its applicability, leaving a wide range of common developer tasks without a suitable abstraction.

\subsection{Refactoring-Aware Code Review}

By recognizing refactorings during the code review process, parts of the textual difference can be consolidated into coherent refactoring operations, thereby increasing the review's efficiency~\cite{coelho2019refactoring}.
Hayashi et al.~\cite{hayashi2013rediffs,zui-apsec2011} proposed an interactive code difference viewer that can separate refactoring changes from other changes.
Ge et al.~\cite{ge2014towards} developed a tool that excludes refactoring-related code changes by reading the refactoring history and displaying the remaining code changes. RefDistiller~\cite{alves2014refdistiller} is a tool checking the existence of missing edits or extra edits when developers manually conduct refactorings to detect potential behavioral changes. Brito et al.~\cite{brito2021raid} propose RAID, which is a refactoring-aware code review tool based on the refactoring detector RefDiff~\cite{silva2017refdiff}. They conducted a field study with eight professional developers, and they found that RAID can reduce the cognitive effort required for reviewing refactorings when using textual diffs. Due to the inherent constraints of refactoring, many code changes are beyond its descriptive reach. Adopting a micro-change-aware approach to code review can further enhance efficiency.

\section{Micro-Changes}

\subsection{Definition}

Based on the need for an intermediary semantic layer within the code modification hierarchy, we present the concept of \textit{micro-change}.

Micro-changes are code change operations described in natural language, designed to bridge the cognitive divide by translating the textual diffs into more understandable natural-language described operations. 

It offers a more efficient and understandable way of examining code changes, encompassing not only refactorings but also a broader spectrum of modifications. The usage of micro-changes aims to redefine how developers interpret and communicate code changes, facilitating a deeper understanding and more efficient conveyance of their essential nature.

An example micro-change (\ReverseCondition) is shown in \cref{f:example_micro_change}; the core change consists only of one line of code (line 363) that inverses the condition of an if statement, while the diff view misleads one to think that there are dozens of changed lines.

Establishing a complete and exhaustive catalog of micro-changes is the wider context of our work, and goes beyond the scope of this paper. Here we focus on a specific category of micro-changes, namely those related to changes pertaining to conditional expressions, such as if statements. 
Those changes are selected because they involve fundamental and essential aspects of logic control in software, which significantly impacts the understanding of behavior through a change. 
Micro-changes are also applicable to different types of changes other than conditional expressions.

\subsection{Catalog of Conditional-Related Micro-Changes}

We have developed a catalog comprising 20 distinct micro-change types, each designed to represent common code change patterns affecting conditional expressions of if statements. 
Due to space constraints within this paper, we provide a concise overview of the catalog in \cref{t:catalog}, which contains only the name and the description of our catalog. The complete version is published for reference\footnote{\url{https://github.com/salab/Micro-Change-Catalog}}.

We also provide an example of a micro-change \AddConjunctOrDisjunct in the catalog shown in \cref{t:example1}.
It contains
\begin{enumerate}
    \item \textit{Name}: a concisely described name in natural language,
    \item \textit{Description}: a description that outlines the nature of the micro-change,
    \item \textit{Structure}: the structure, explaining the micro-change in pseudo-code,
    \item \textit{Motivation}: a motivation for the existence of this micro-change,
    \item \textit{Example}: one or more examples taken from open-source repositories, to illustrate the micro-change in practice, and
    \item \textit{Detection Rule}: a specific detection rule designed to identify this micro-change in the wild.
\end{enumerate}

\begin{table}[ht]
    \centering
    \caption{Catalog Entry for \AddConjunctOrDisjunct}
    \label{t:example1}
    \begin{tabular}{p{8.4cm}}
        \toprule
        \textbf{\AddConjunctOrDisjunct}\\ \midrule
        \textbf{Description}\\
        Modify an existing conditional statement by appending an additional condition using logical AND (\texttt{\&\&}) or logical OR (\texttt{||}) operators to refine or expand the criteria for the statement's execution. \\ \midrule

        \textbf{Structure}
        \begin{itemize}
            \item $\mathtt{if}~(\$c_1) \rightarrow \mathtt{if}~(\$c_1~\text{\texttt{\&\&}}~\$c_2)$
            \item $\mathtt{if}~(\$c_1) \rightarrow \mathtt{if}~(\$c_1~\text{\texttt{||}}~\$c_2)$
        \end{itemize}
        where $\mathtt{if}~(\cdots)$ represents an if statement, $c_1$ and $c_2$ denotes two conditional expressions. The $\text{\texttt{\&\&}}$ and $\text{\texttt{||}}$ denote the logical AND and logical OR operators, respectively.\\ \midrule

        \textbf{Motivation}\\
        Attach extra conditions to refine or expand the criteria for the statement's execution with logical AND ($\text{\texttt{\&\&}}$) or logical OR ($\text{\texttt{||}}$) operators, respectively.\\ \midrule

        \textbf{Example}\\
        An example of \AddConjunctOrDisjunct found in the repository~\repository{zuul}\footnote{https://github.com/Netflix/zuul/commit/1a47cc0}:\\
        \includegraphics[width=\linewidth]{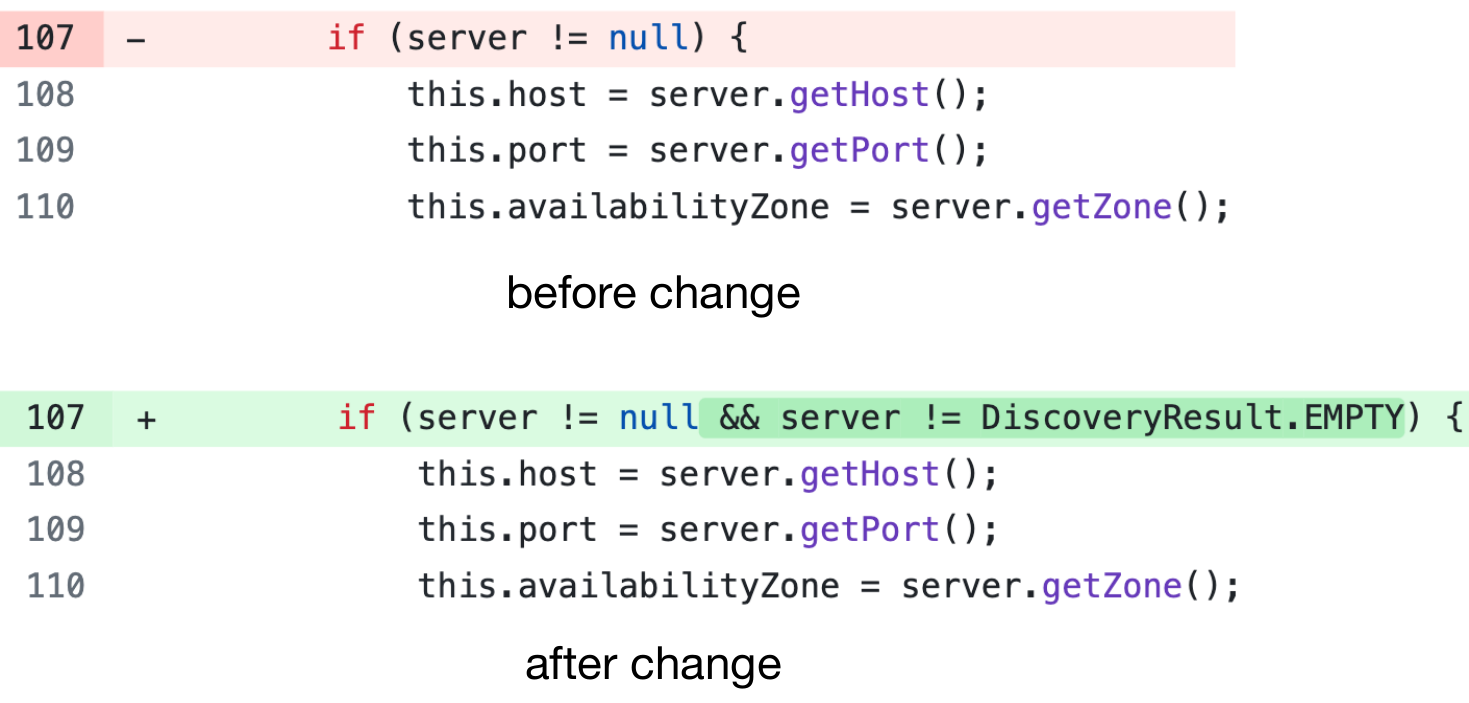}

        The new condition \texttt{server != DiscoveryResult.EMPTY} is attached to the \texttt{server != null} to make the condition more strict by requiring an extra criterion to be met.\\ \midrule

        \textbf{Detection Rule}\\
        \textit{Detection rules are introduced in \cref{s:detecting_micro-change}.}\\
        \bottomrule
    \end{tabular}
\end{table}

The construction of the catalog went hand-in-hand with the detection of micro-changes, following a multi-step iterative process, which we explain in \cref{s:detecting_micro-change}.

\subsection{Research Questions} \label{ss:research_questions}

To have a deeper understanding of the proposed micro-changes and the catalog built, we propose two research questions as follows.

\def\RQOne{To what extent can the designed micro-changes explain the code change?}
\def\RQTwo{What are the frequencies of the micro-changes types?}

\begin{description}

\item [\RQ{1}:] \textit{\RQOne}
The purpose of \RQ{1} emerges from the necessity to assess the comprehensiveness and completeness of the developed conditional-related micro-change catalog. It can also uncover opportunities for future enhancements.

\item [\RQ{2}:] \textit{\RQTwo}
By quantifying how frequently each micro-change type occurs, \RQ{2} investigates their prevalence and distribution across repositories. Furthermore, identifying the frequency of various micro-changes can help prioritize which types to focus on for further refinement in the catalog and enhancement of detection algorithms.

\end{description}

\begin{figure*}[ht]\centering
    \includegraphics[width=\linewidth]{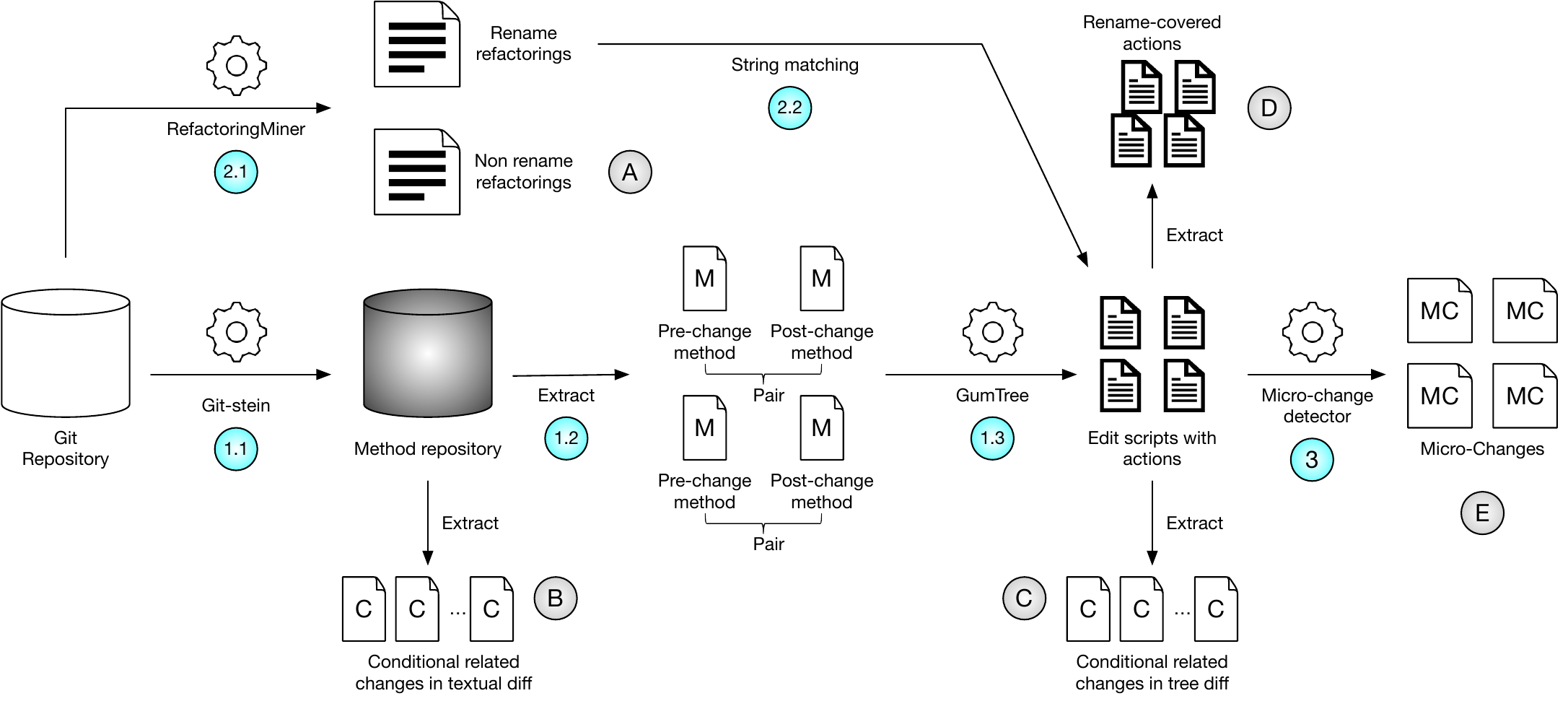}
    \caption{Overview of micro-change detection.}
    \label{f:micro-change-detecting}
\end{figure*}

\section{Detecting Micro-changes} \label{s:detecting_micro-change}

To validate the catalog of micro-changes, ensure its completeness, and further investigate the features of micro-changes, we designed a process for detecting the micro-changes in the catalog in the wild.

\subsection{Detection Process}
The overview of the micro-change detection process is shown in \cref{f:micro-change-detecting}. Starting from the initial input, a Git repository is transformed into a method-level repository, where pairs of pre-change and post-change methods are extracted. These pairs undergo analysis by a tree-diff tool to generate \textit{edit scripts} and \textit{actions}. Concurrently, refactorings are extracted from the Git repository, within which the \refactoring{Rename}-related refactorings are used to identify the actions that can be annotated as renaming. By excluding these renaming actions, the micro-change detector detects micro-changes in the remaining. The subsequent sections detail each step of this process:

\paragraph*{\textbf{1.1: Derive Method-Level Repository}} The Git repository is converted to a method-level repository, where each method in the original repository is extracted into an individual file. This enables Git to track changes at method level~\cite{higo2020tracking, hata2011historage}, which is the level of our interest since conditional expressions are exclusively found in methods. For the conversion of the repository, we use the tool \textit{git-stein}\footnote{\url{https://github.com/sh5i/git-stein}}\cite{shiba-jssst202211}. 

\paragraph*{\textbf{1.2: Extract Changed Methods}}\label{ss:extract_changed_methods} From the method repository we obtained from the last step, a complete history of each method can be extracted. Within this history, methods before and after each commit are identified as method pairs representing the changes. We extract all such pairs from the method history to serve as input for the subsequent analysis step, excluding instances where methods are either newly introduced or entirely deleted. Note that method renaming is treated as a special case of deletion and introduction; the original method is considered deleted, and a new method with the new name is introduced, thereby excluding these renaming events from our analysis of method changes.

\paragraph*{\textbf{1.3: Extract AST Edit Scripts}} We use GumTree~\cite{falleri2014fine} to analyze the pairs of pre-change and post-change methods and extract the edit scripts. The GumTree supports only several programming languages. The idea of micro-change is language-agnostic and can be further extended to be applied to other programming languages by substituting the edit script extraction tool and corresponding predefined detection rules.

An edit script consists of a series of actions, such as \emph{insert}, \emph{delete}, \emph{update}, and \emph{move}, detailing how to transform the original AST into the modified version.

The \emph{insert} action identifies a new AST node that has been added to the AST, represented as $\INS(\mathit{node}, \mathit{parentNode})$, where the $\mathit{node}$ specifies the newly added AST node and the $\mathit{parentNode}$ refers to the node under which the new node is inserted. The \emph{delete} action detects an AST node present in the original AST that is absent in the modified version of AST and denoted as $\DEL(\mathit{node})$, where $\mathit{node}$ is removed. The \emph{update} action represents a node in the AST that has been altered to the same type but a different value, such as modifying a variable name or changing the literal value. It is denoted as $\UPD(\mathit{node}, \mathit{val})$, where $\mathit{node}$ is the node being altered and $\mathit{val}$ is the new value assigned to $\mathit{node}$. The \emph{move} action signifies that an AST node be relocated within the AST, maintaining its original structure but changing its location in the code. It is denoted as $\MOV(\mathit{node}, \mathit{parentNode})$, where $\mathit{node}$ is the moved node, $\mathit{parentNode}$ is the node above the moved node at the post-change AST.

We exclude the edit scripts consisting of only insertions or only deletions, as they would be part of pure addition or pure deletion commits, and not change-related commits.

\paragraph*{\textbf{2.1: Extract Refactorings}} Starting from the original Git repository, we use RefactoringMiner 3.0.4~\cite{Tsantalis:TSE:2020:RefactoringMiner2.0} to extract all refactorings performed during its history. However, while it is capable of identifying where refactorings have been applied, it is currently not capable of revealing which other locations in the source code are impacted by refactorings, which is necessary in the specific case of rename refactorings. 

\begin{figure}[ht]\centering
    \includegraphics[width=\linewidth]{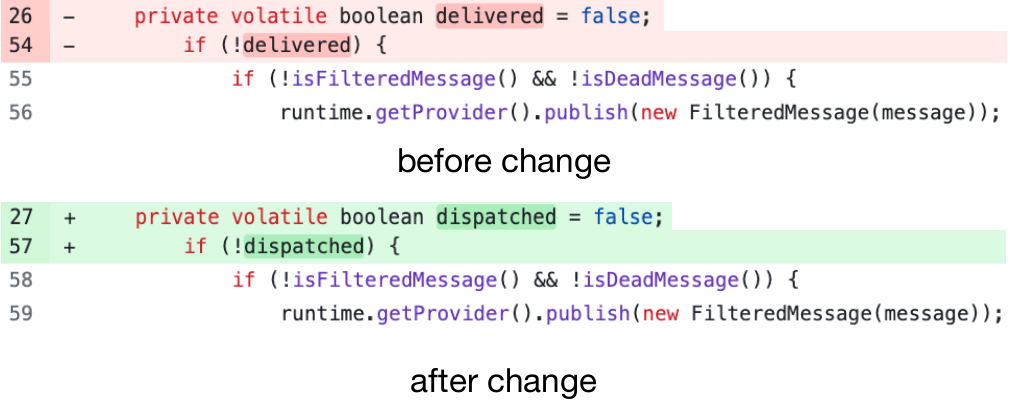}
    \caption{\refactoring{Rename Attribute} refactoring.}
    \label{f:rename_attribute}
\end{figure}

To illustrate this point, let us consider the example shown in \cref{f:rename_attribute}.
It is a commit from the repository \repository{mbassador}\footnote{\url{https://github.com/bennidi/mbassador/commit/744c029}}, where a \refactoring{Rename Attribute} refactoring is detected; \attribute{delivered} is renamed to \attribute{dispatched}, transitioning from line 26 to line 27 post-change. This renaming influences the conditional expressions previously at line 54 and modified to line 57. Despite these changes being a direct result of the refactoring, RefactoringMiner does not identify such downstream effects.

To address the identified issue, we enhanced the functionality of RefactoringMiner, to also output the code locations affected by a refactoring; if an \emph{update} action showing that value $Y$ is assigned to a code element whose original value is $X$, and a \refactoring{Rename}-related refactoring is detected within the same file at the same commit that renames $X$ to $Y$, we classify this action as a downstream effect of the refactoring.

\paragraph*{\textbf{2.2: Apply String Matching}} In commit where the \refactoring{Rename}-related refactorings are detected, we conduct a string-matching; if the action in the edit script within a commit is updating the name of a code element, and it belongs to the downstream effect of the \refactoring{Rename}-related refactoring, we regard this action as covered by the \refactoring{Rename} refactoring. Such actions are excluded from the input considered by the micro-change detector because refactoring constitutes a specific category of micro-changes.

\paragraph*{\textbf{3: Detect Micro-Changes}}
To detect micro-changes of conditional-related changes, we rely on a set of predefined rules coupled with the analytical capabilities of the tree diff technique. These predefined rules are designed to identify specific patterns that pertain to micro-changes, facilitating a targeted and efficient analysis.

The detection rule of the micro-change \AddConjunctOrDisjunct is:
\begin{equation}
    \mathrm{belongs}(e, \mathit{cond}) \wedge \INS(\mathit{IEO}.\&\&, e) \vee \INS(\mathit{IEO}.||, e)
\end{equation}
where $e$ and $\mathit{cond}$ are AST nodes that represent expression and conditional expression in an if statement, respectively. The $e$, a child node of $\mathit{cond}$, is represented as $\mathrm{belongs}(e, \mathit{cond})$. The $\mathit{IEO}.\&\&$ and $\mathit{IEO}.||$ represent the infix expression operator, logical AND ($\&\&$), and infix expression operator, logical OR ($||$), respectively. The $\INS(\mathit{IEO}.\&\&, e)$ indicates an insertion action of inserting the logical AND to the expression $e$. In other words, the rule can be expressed as within an existing conditional expression in an if statement, the insertion of a logical AND or logical OR operator is considered indicative of the \AddConjunctOrDisjunct micro-change. This criterion is used to identify instances where an additional condition is appended to an existing conditional expression.

\subsection{Catalog Creation}

Our objective was to develop a comprehensive catalog of micro-changes targeting conditional-related changes, capturing a spectrum of these changes as broad as possible. Recognizing the significant time investment and effort required to build an exhaustive catalog, we adopted a systematic approach to manage the task within a feasible timeframe. 

We began with the preliminary identification of 13 distinct micro-change types, characterized by the addition, deletion, and alteration of code elements within conditional expressions.

Then, to refine and expand our initial categorization, we selected the \repository{mbassador} repository\footnote{\url{https://github.com/bennidi/mbassador}}, which contains 226 conditional-related changes. The first author undertook three review cycles, focusing on changes not covered by our defined micro-change types.

During the first review, we discovered that 166 out of the total conditional changes fell outside of our catalog's scope.

We randomly picked a subset of 100 changes as a basis for enhancing the existing micro-change designs and for formulating additional new micro-change types. Following this refinement and a subsequent detection phase, the second review revealed 90 changes still not accommodated by our updated catalog. This prompted further enhancements and the introduction of new micro-change types, culminating in a catalog comprising 20 distinct types.

Upon implementing these revisions, a third review round was carried out, revealing that out of the 226 conditional-related change, 64 were still not covered by our micro-change catalog. However, these instances can all be attributed to incorrect parsing by GumTree, the tool we utilized for deriving tree diff. It occasionally fails to recognize the holistic structure of certain code modifications. For example, GumTree should recognize a change as removing a tree, where the root node is a conditional expression, and its children represent the associated code block. However, it occasionally misinterprets the change as removing a series of individual nodes, rather than recognizing the entire tree structure as a single conditional block removal.

Ultimately, our refined catalog achieved a 71.7\% coverage rate of conditional-related changes, demonstrating substantial progress toward our goal of comprehensive coverage.

\begin{table*}[ht]\centering
    \caption{Overview of Dataset Comprising 73 Projects}
    \label{t:overview_of_dataset}
    \rowcolors{2}{gray!10}{white}
    {
    \begin{tabular}{lrrrrrr} \toprule
        \textbf{Metric} & \textbf{Min} & \textbf{Q1} & \textbf{Median} & \textbf{Q3} & \textbf{Max} & \textbf{Total} \\ \midrule
        Java files & 39 & 425 & 814 & 2,067 &  10,012 & 114,080 \\
        Lines of code & 4,200 & 33,681 & 64,048 & 166,517 &  970,310 & 10,931,743 \\
        Commits    & 342 & 2,025 & 3,588 & 7,810 &  14,971 & 366,697 \\
        Processed commits & 254 & 1,732 & 3,298 & 6,179 &  14,605 & 312,554 \\
        Processed methods & 208 & 7,341 & 17,143 & 40,085 &  178,437 & 2,051,128 \\
        Conditional-related commits & 17 & 818 & 2,245 & 5,118 &  11,753 & 240,201 \\
        Conditional-related changes & 36 & 1,925 & 5,609 & 13,504 &  33,222 & 637,669 \\ \bottomrule
  \end{tabular}}
\end{table*}

\section{Empirical Study}

\subsection{Overview}

To investigate micro-changes with respect to helping understand code changes, and to answer the research questions proposed in \cref{ss:research_questions}, we conducted micro-change detection on a large-scale open-source Java repositories dataset.

To answer \RQ{1}, we calculate the coverage of micro-changes on conditional-related changes. We regard changes to if statements, their direct children, and any descendants of their conditional expressions as \emph{conditional-related changes}.
For \RQ{2}, the frequency of the micro-changes is calculated to uncover which type of micro-changes are commonly used.

\subsection{Data Collection}

The dataset we used is the one collected by Silva et al.~\cite{silva2016we}, which contains 124 GitHub-hosted software projects. 
The selection of repositories for the dataset was governed by a set of criteria to ensure the relevance and feasibility of our analysis. Firstly, because the total number of commits of the repository is not a key feature influencing our investigation on micro-changes, we imposed a threshold on the number of commits, excluding any repositories with more than 15,000 commits to manage the complexity and execution time of our experiments. Secondly, we required that the repositories remain accessible via the links provided in the dataset curated by Silva et al., ensuring that our sources were both current and retrievable. Last, we excluded non-Java projects, as our current detection mechanism is specialized on Java source code. Employing these criteria, we composed a dataset comprising 73 repositories.
The dataset is available in our supplementary package\cite{dataset}.

\begin{figure}[b]\centering
    \includegraphics[width=1.0\linewidth]{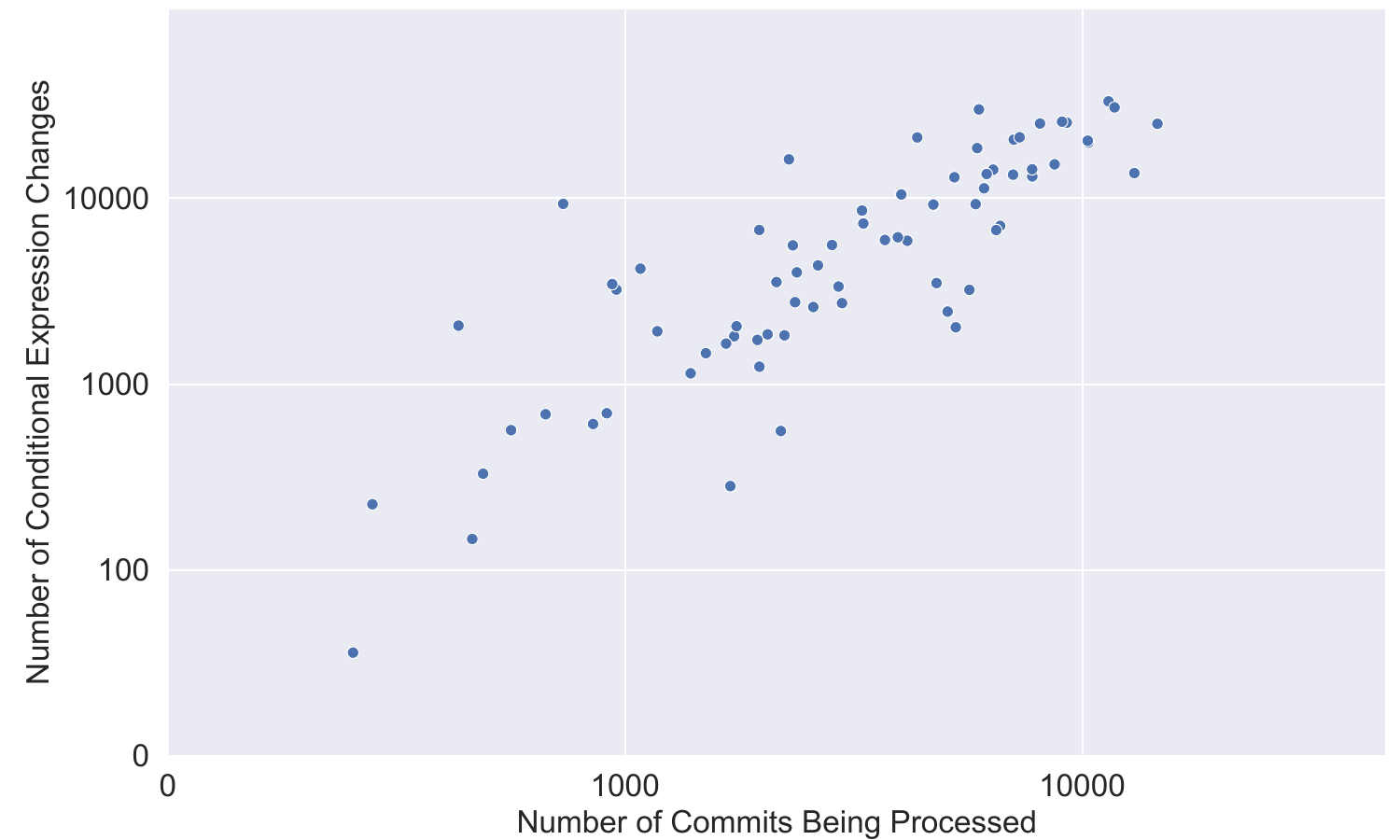}
    \caption{Distribution of dataset.}
    \label{f:dataset_info}
\end{figure}

A plot of our dataset is shown in \cref{f:dataset_info}, where the x-axis is the number of commits being processed in our experiment, and the y-axis is the number of conditional-related changes. Basically, the number of conditional-related changes increases with the number of commits being processed.

We also show an overview of our dataset in \cref{t:overview_of_dataset}, where the term \textit{Processed commits} denotes the count of commits after excluding merge commits to avoid duplicate detection, excluding commits involving only additions or only deletions which is not the scope of this study. Similarly, the \textit{Processed methods} pertains to the count of methods after removing those that have exclusively additions or deletions.
In total, we processed 366,697 commits, where 240,201 of them feature changes affecting conditional expressions. The total number of conditional-related changes that we analyzed is 637,669.

\subsection{\RQ{1}: \RQOne}

\def\Ref{\mathit{Ref}}
\def\Rename{\mathrm{rename}}
\def\NonRename{\text{non-rename}}
\def\CRC{\mathit{CRC}}
\def\Diff{\mathit{Diff}}
\def\Text{\mathrm{text}}
\def\Tree{\mathrm{tree}}
\def\MC{\mathit{MC}}
\def\Coverage{\mathit{Coverage}}
\def\CoverageMC{\Coverage^\text{MC}}
\def\CoverageMCRM{\Coverage}
\def\CoverageRM{\Coverage^\text{RM}}

\subsubsection{Study Design}

This research question aims to evaluate the breadth and depth with which our catalog of designed micro-changes can capture and accurately represent the essence of code changes. We implemented the micro-change detection methodology outlined in \cref{s:detecting_micro-change} across our dataset to facilitate this analysis.

To quantitatively measure the outcome, we developed a metric measuring the coverage of conditional-related changes encompassed by our catalog of micro-changes.

As illustrated in \cref{f:micro-change-detecting}, the annotations \Circled{A} and \Circled{D} denote lines of code changes identified as non-rename refactoring ($\Ref^\NonRename$) and rename refactoring ($\Ref^\Rename$), respectively. Their union can capture the code changes involved with any refactorings detected by RefactoringMiner:
\begin{equation}
    \Ref = \Ref^\NonRename \cup \Ref^\Rename.
\end{equation}
Annotations \Circled{B} and \Circled{C} indicate the conditional-related changes identified through textual-diff ($\Diff^\Text$) and tree-diff ($\Diff^\Tree$), respectively, while \Circled{E} highlights those condition-related changes captured by detected micro-changes in the catalog ($\MC$).

The collective set of lines comprising conditional-related changes emerges from the intersection of $\Diff^\Text$ and $\Diff^\Tree$. This intersection leverages the strengths of both tree-based and textual diff algorithms to heighten the precision in pinpointing code alterations, mitigating the risk of falsely identifying unchanged code as changed. By integrating these approaches, the likelihood of false positives is reduced, thereby increasing precision in the detection of genuine code changes. The set of conditional-related changes ($\CRC$) is denoted as:
\begin{equation}
    \CRC = \Diff^\Text \cap \Diff^\Tree.
\end{equation}
A subset of these conditional-related changes are addressed by our micro-changes.

The ratio quantifying micro-change coverage together with the detected refactorings ($\CoverageMCRM$) is computed as:
\begin{equation}
    \CoverageMCRM = \frac{|(\MC \cup \Ref) \cap \CRC|}{|\CRC|}.
\end{equation}

To underscore the unique contributions of micro-changes and traditional refactorings in capturing code changes, we introduce two distinct coverage ratios.

The \emph{micro-change coverage} ($\CoverageMC$) focuses solely on the extent to which micro-changes from our catalog cover conditional-related changes, reflecting the specific impact of our catalog's micro-changes, which is denoted as:
\begin{equation}
    \CoverageMC = \frac{|\MC \cap \CRC|}{|\CRC|}.
\end{equation}

The \emph{refactoring coverage} ($\CoverageRM$) quantifies the extent of Rename and Non-Rename Refactorings identified by RefactoringMiner, showcasing the impact of traditional refactorings, which is denoted as:
\begin{equation}
    \CoverageRM = \frac{|\Ref \cap \CRC|}{|\CRC|}.
\end{equation}

These metrics are crucial for clearly distinguishing the contributions of micro-changes from refactorings, ensuring a precise assessment of their roles in software development practices without overlap. The results of $\CoverageMCRM$, $\CoverageMC$, and $\CoverageRM$ are compared to prove the effectiveness of our micro-change catalog in expressing code changes.

In addition, to validate the credibility of the coverage result, we randomly selected 100 detected micro-change instances from four repositories within the dataset. The first author of this paper manually reviewed those instances.

\subsubsection{Results and Discussion}

\begin{figure}[ht]
    \centering
    \includegraphics[width=1.0\linewidth]{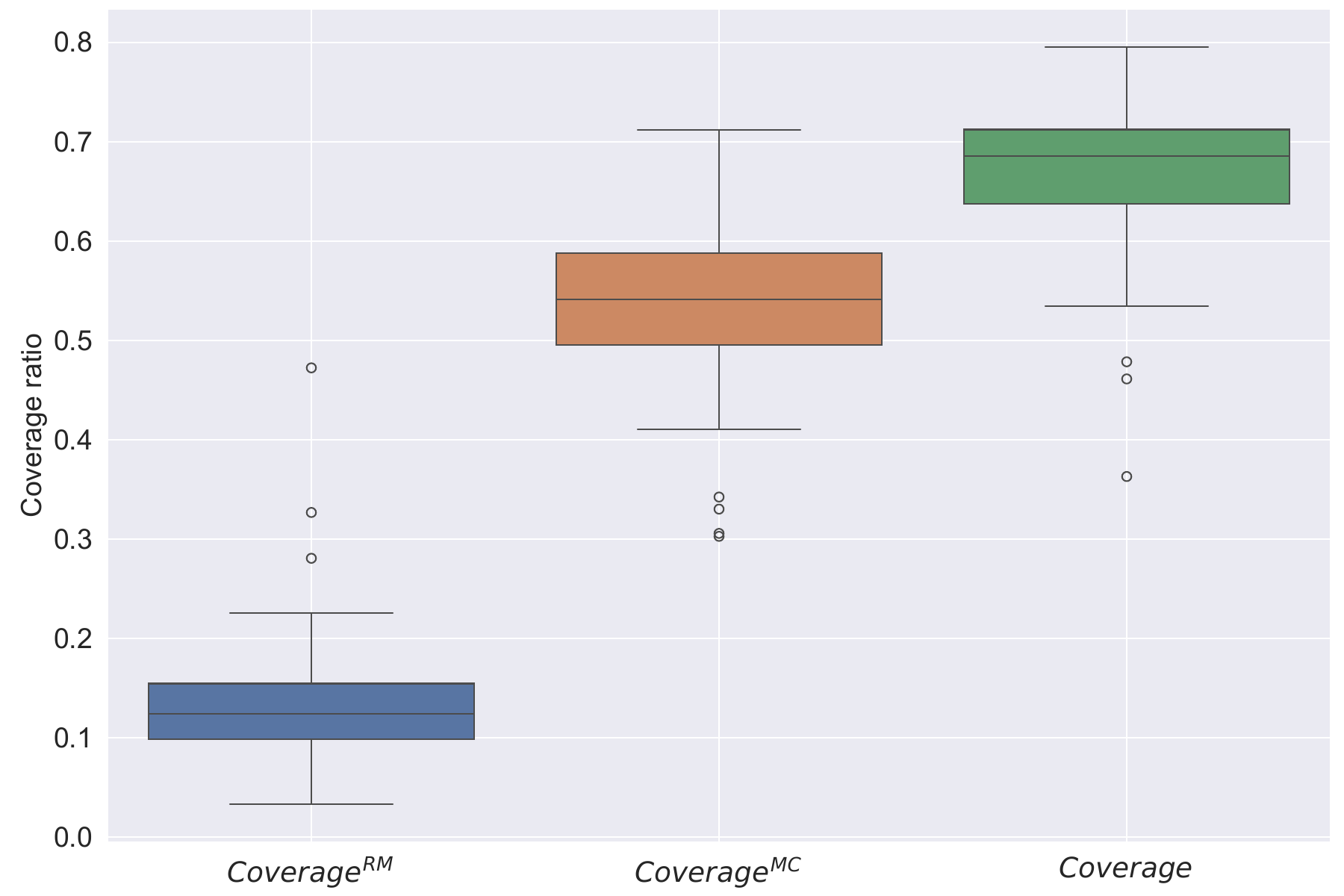}
    \caption{Coverage across the dataset.}
    \label{f:coverage_box_plot}
\end{figure}

\begin{figure*}[ht]\centering
  \includegraphics[width=\linewidth]{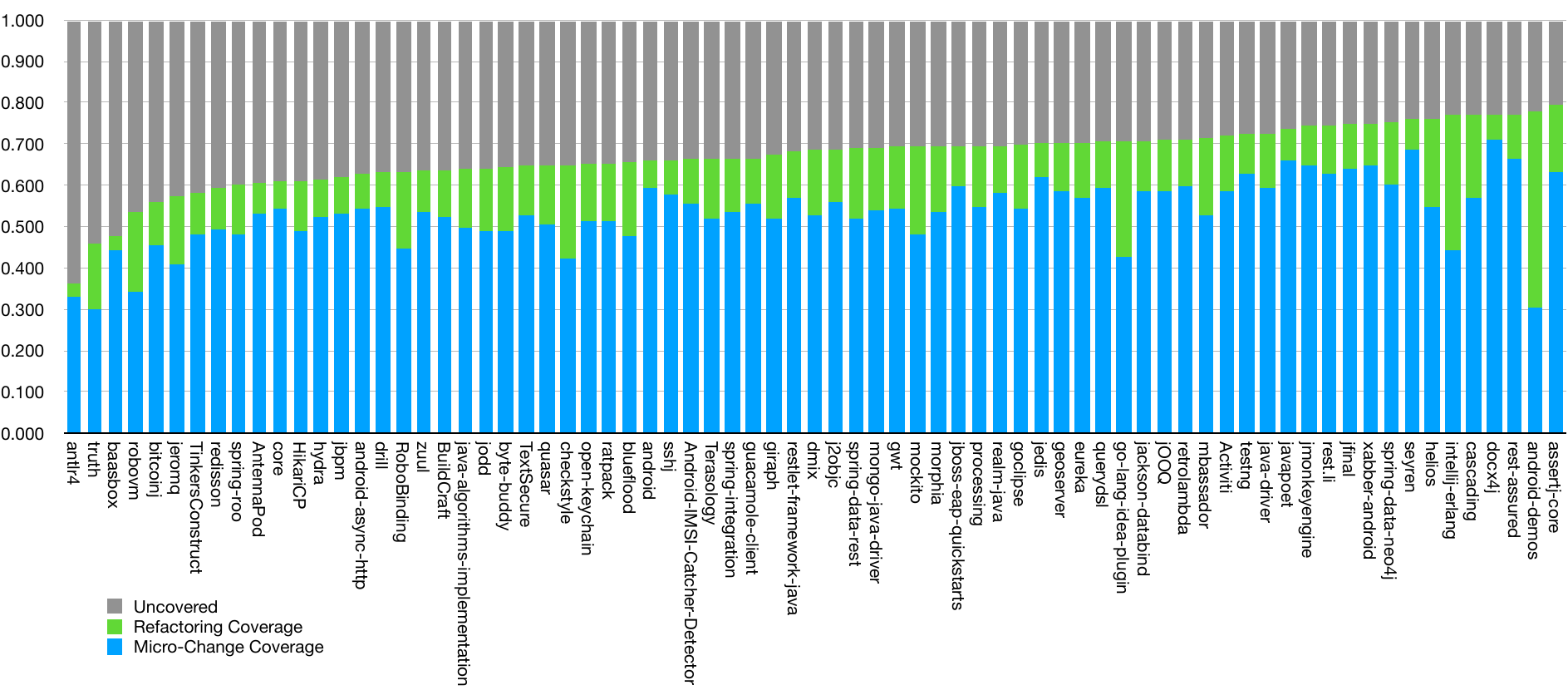}
  \caption{Coverage across all 73 repositories, sorted according to the sum of Refactoring coverage and Micro-Change Coverage.}
  \label{f:coverage_all_73}
\end{figure*}

In general, the coverage of micro-changes on our dataset has an average ratio of 67.1\%. The distribution is depicted in the boxplot in \cref{f:coverage_box_plot}.
The three boxes correspond to the $\CoverageRM$, $\CoverageMC$, and $\CoverageMCRM$, indicating the coverage of RefactoringMiner, the coverage of micro-changes, and the combined impact of both. The medians for the three boxes are 0.124, 0.541, and 0.685, while the average values are 0.135, 0.536, and 0.671.

A comparative analysis between the $\CoverageRM$ and $\CoverageMC$ reveals a significant insight: There is a significant difference between these two, as evidenced by a paired t-test ($p<0.001$) and Cohen's $d$ of 5.5, which strongly suggests the difference is not by random chance. Our micro-change catalog achieves, on average, four times greater coverage of conditional-related changes than refactorings alone. This finding underscores the substantial efficacy of our catalog in capturing a broader spectrum of code changes, thereby affirming its value and precision. The coverage across each repository within our dataset of 73 repositories is detailed in \cref{f:coverage_all_73}.

With the exception of one repository~\repository{android-demo}, the remaining 72 repositories exhibit, on average, ca. 5 times greater $\CoverageMC$ than $\CoverageRM$. Notably, the repository~\repository{bassbox} has a $\CoverageMC$ of 0.444, which is 13 times greater than $\CoverageRM$, standing at 0.034.

Furthermore, an analysis reveals that 67.1\% of the repositories possess $\CoverageRM$ of less than 0.15. This low coverage suggests that refactorings alone are insufficient in fully accounting for the essence of code changes within conditional expressions. In contrast, the average micro-change coverage ($\CoverageMC$) across these repositories stands at 0.56, which is approximately 3.7 times greater. This disparity indicates that the coverage achieved by combining both refactorings and micro-changes ($\CoverageMCRM$) is largely attributed to the micro-changes that we have cataloged. The micro-changes in our catalog thus serve as a complementary extension to refactorings, significantly enhancing the ability to explain conditional-related changes.

To ensure the correctness, the first author manually reviewed 100 detected micro-changes randomly extracted from four repositories. The review revealed that 86\% of these instances were true positives, underscoring the reliability of our detection methodology.

The causes of false positives can be divided into two: 1) inaccuracies in diff parsing by GumTree, and 2) instances where a single change undergoes multiple micro-changes, making it impossible to capture a change accurately with just one micro-change.

The first issue underscores a clear opportunity for enhancement within our detection methodology. 
Employing a more precise tree-diff tool could markedly increase the detection accuracy, pointing to a vital area for technical refinement.

\begin{figure}[t]
    \centering
    \includegraphics[width=1.0\linewidth]{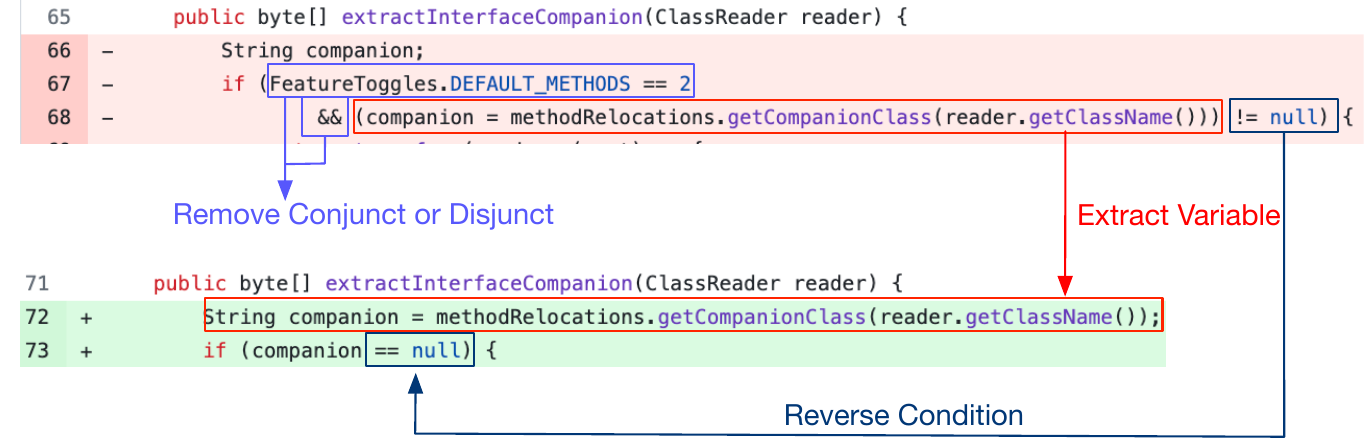}
    \caption{Multiple micro-changes in a single conditional expression.}
    \label{f:compound_micro-change}
\end{figure}

The second issue introduces the concept of \textit{compound micro-change}, where multiple micro-changes are applied to a single code element.
An example is shown in \cref{f:compound_micro-change}.
It is a code change sourced from the repository~\repository{retrolambda}\footnote{\url{https://github.com/luontola/retrolambda/commit/e3c8647}}.
Here, the code modification involves three distinct micro-changes: \RemoveConjunctOrDisjunct, \microchange{Extract Variable} and \ReverseCondition.
This change is too complicated for tree diff to comprehensively parse the sequence of actions.
As a result, only the \RemoveConjunctOrDisjunct is detected, with the additional code being misclassified as a new insertion (\textit{Add Conditional Statement}) due to its complexity. 

Additionally, we examined changes in conditional expressions that fell outside the scope of both our micro-change catalog and refactorings. 
The same two issues that contributed to false positives were also responsible for these uncovered changes. 
Furthermore, though we did not observe in our review process, we do not deny the possibility that certain changes possess the potential to be classified as new types of micro-changes, indicating an opportunity for expanding our catalog to encompass a broader spectrum of modifications.

\Conclusion{On average, micro-changes can explain 67.1\% of the conditional-related changes.
Micro-changes in the catalog have a better ability to explain changes than refactorings. The detection mechanism achieves an accuracy of 86\%.}

\subsection{\RQ{2}: \RQTwo}

\subsubsection{Study Design}

In this research question, we calculate the detected frequency of the micro-changes designed in our catalog in the dataset. In addition, we also deduce the developers' motivation for introducing the most frequent micro-change type in the development.

The set of micro-changes detected from the dataset of type $t \in T$ is denoted as $MC_t$, where $T$ is the set of the types in our catalog. The frequency of a micro-change of type $t$ is calculated as:
\begin{equation}
    \mathit{freq}(t) = \frac{|MC_t|}{\sum_{t'\in T}{|MC_{t'}|}}.
\end{equation}

\subsubsection{Results and Discussion}

\begin{table}[ht]\centering
    \caption{Frequency of Micro-changes Detected}
    \label{t:frequency_of_micro_changes}
    \rowcolors{2}{gray!10}{white}
    {\tabcolsep=5pt\begin{tabular}{lrrr} \toprule
        \textbf{Micro-change} & \hspace{-3em}\textbf{Occurrences} & \textbf{Frequency} & \textbf{Rank} \\ \midrule
        \AddConditionalStatement & 2,931,551 & 28.10\% & 1\\
        \WrapStatementInBlock & 2,569,619 & 24.60\% & 2\\
        \RemoveConditionalStatement & 1,245,424 & 11.90\% & 3\\
        \UnwrapStatementFromConditional & 954,654 & 9.20\% & 4\\
        \WrapStatementInConditional & 919,582 & 8.80\% & 5\\
        \RemoveConjunctOrDisjunct & 501,002 & 4.80\% & 6\\
        \ReverseCondition & 362,179 & 3.50\% & 7\\
        \AddConjunctOrDisjunct & 316,263 & 3.00\% & 8\\
        \ExtendElseWithIf & 208,262 & 2.00\% & 9\\
        \UnwrapStatementFromBlock & 142,066 & 1.40\% & 10\\
        \MoveInwardCondition & 56,305 & 0.50\% & 11\\
        \RemoveElse & 42,201 & 0.40\% & 12\\
        \ConditionalToBooleanReturn & 38,046 & 0.40\% & 13\\
        \ExtendIfWithElse & 37,886 & 0.40\% & 14\\
        \FlipLogicOperator & 29,949 & 0.30\% & 15\\
        \ConditionalToExpression & 25,475 & 0.20\% & 16\\
        \MoveOutwardCondition & 19,809 & 0.20\% & 17\\
        \ConditionalToSwitch & 18,640 & 0.20\% & 18\\
        \SwapThenAndElse & 8,167 & 0.10\% & 19\\
        \AdjustConditionBoundary & 6,004 & 0.10\% & 20\\ \midrule
        \textbf{Total} & \hspace{-3em}\textbf{10,433,084} & & \\ \bottomrule
  \end{tabular}}
\end{table}

The frequencies of each type of micro-changes are detailed in \cref{t:frequency_of_micro_changes}.
For a given micro-change type $t$, the table lists the number of detected instances in the dataset, the $\mathit{freq}(t)$, along with their respective rank.

The highest frequency micro-change type is \AddConditionalStatement, exhibiting a frequency of 28.1\%, whereas the \AdjustConditionBoundary is the least frequent, with a frequency of 0.1\%. The second most frequent type is \WrapStatementInBlock, indicating a prevalent trend towards discarding syntactic sugar in Java, which allows omitting curly braces surrounding a single statement at a then/else clause. The third most frequent type, \RemoveConditionalStatement, in conjunction with the most frequent \AddConditionalStatement, implies that modifications involving the introduction or removal of entire conditional blocks are more common than minor adjustments, such as changing the logic operators or boundary conditions.
Notably, both changes are {\em not} behavior-preserving, i.e., they represent actual changes in the program logic.

Furthermore, the \AddConjunctOrDisjunct and \RemoveConjunctOrDisjunct occupy the 8th and 6th ranks, respectively. This indicates a relatively high occurrence of refinements in conditional expressions, either by incorporating new conditions or simplifying them. Conversely, the micro-change types \ConditionalToBooleanReturn, \ConditionalToSwitch, and \ConditionalToExpression rank 13th, 18th and 16th, respectively, indicating their low prevalence. This could be attributed to the less frequent application of \textit{Boolean Return}, \textit{Switch}, and \textit{Conditional Operator} compared to other more common patterns.

For the most frequent micro-change type \AddConditionalStatement, we categorized the motivations of developers to introduce this micro-change into three types:

\begin{enumerate}

\item \textbf{Security/Compatibility Check}: This type of conditional statement addition emerges from ensuring that only users or systems meet specific security criteria or have compatible resources that can execute the operations. An example is found in repository~\repository{retrolambda}\footnote{\url{https://github.com/luontola/retrolambda/commit/2501461}} and illustrated in \cref{f:security_check}.
The added code block is to ensure the program only operates on Java 8, preventing runtime errors or unexpected behavior on incompatible Java versions.

\begin{figure}[bh]\centering
  \includegraphics[width=1.0\linewidth]{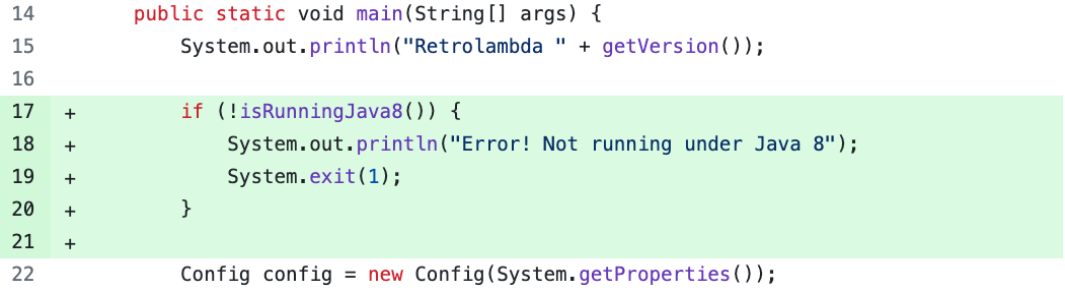}
  \caption{Conditional statement added for security check.}
  \label{f:security_check}
\end{figure}

\item \textbf{Performance Optimization}: The motivation for this type is more about performance, which is to check for and utilize optimal configurations or system conditions that enhance the program's performance. Depicted in \cref{f:optimize_performance}, the example found in repository~\repository{retrolambda}\footnote{\url{https://github.com/luontola/retrolambda/commit/281ef93}}, as commented by the developer, the code returns earlier on classes whose names start with \textit{``java/''} to avoid continue processing the unnecessary operations.

\begin{figure}[ht]\centering
  \includegraphics[width=1.0\linewidth]{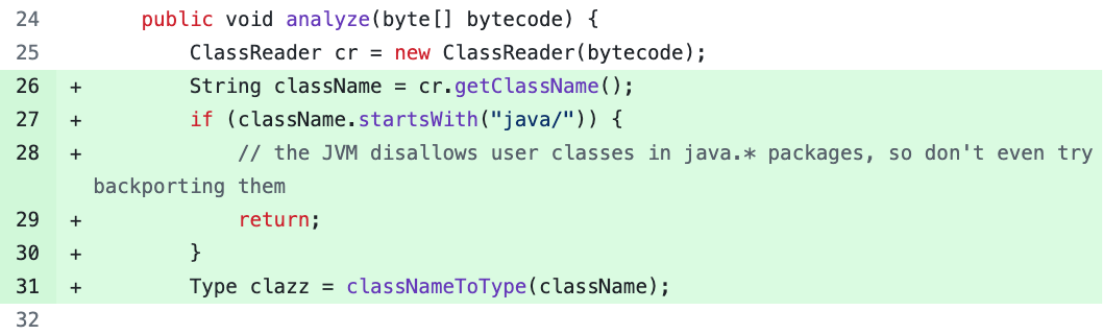}
  \caption{Conditional statement added for performance optimization.}
  \label{f:optimize_performance}
\end{figure}

\item \textbf{User Assistance}: This type of insertion will not affect the functionality of the program, and it is to provide feedback or help to the user if the program's prerequisites are not met, enhancing user experience or improving the user's ability to resolve configuration issues more efficiently.
An example is found in repository~\repository{zuul}\footnote{\url{https://github.com/Netflix/zuul/commit/040d3ef}} and shown in \cref{f:user_assistance}.
By checking the format of the input and logging an error if it does not meet the expected format, the system provides immediate feedback that can help the user or developer identify and correct the issue.

\begin{figure}[h]\centering
  \includegraphics[width=1.0\linewidth]{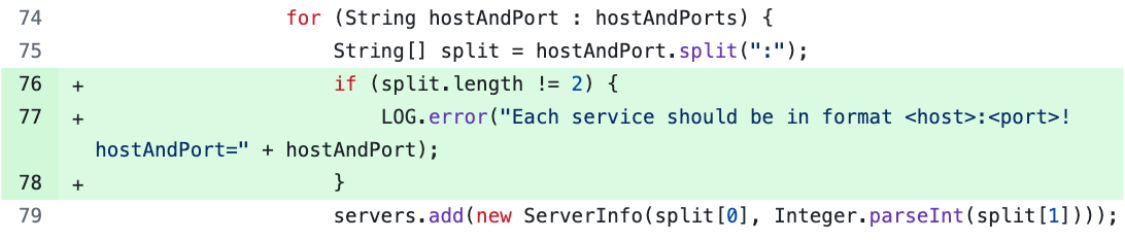}
  \caption{Conditional statement added for user assistance.}
  \label{f:user_assistance}
\end{figure}

\end{enumerate}

\Conclusion{We reported the frequency of each type of designed micro-change in the catalog. The most frequent types are \AddConditionalStatement, \WrapStatementInBlock, and \RemoveConditionalStatement, which together account for nearly 2/3 of the micro-changes.}

\section{Threats to Validity}

\paragraph*{Internal Validity} The concept of micro-change is defined broadly and abstractly, which also enables us to define various trivial and uninteresting types, according to the definition of the micro-change, e.g., addition or removal of a specific character. We leave open the discussion about the boundary between micro-changes and non-micro-changes. Indeed, there is still a lot of beneficial vocabulary to describe changes that are not covered by the well-known existing change vocabulary, such as refactoring operations. Another possible threat is that we excluded the renaming of methods as introduced in \cref{ss:extract_changed_methods}. We could treat the renamed method as the same file and examine its differences, rather than just the renaming itself. While this is feasible, it adds complexity. To simplify our analysis, we chose to ignore renaming.

\paragraph*{Construct Validity} The comprehensiveness of our conditional-related micro-change catalog is critical for the validity of our findings regarding the scope of code change it can explain. There is a possibility that our catalog may not encompass all relevant micro-change types, especially given the vast and evolving nature of software development practices. As a result, our analysis and conclusions about how common and important different types of micro-changes might not be completely accurate. However, adding more micro-change types to the catalog does not impact the existing ones, but rather lowers even more the changes unaccounted for.

\paragraph*{External Validity} Our selection criteria for repositories, specifically those with fewer than 15,000 commits, may limit the generalizability of our findings. This threshold was chosen to manage the scale and complexity of the analysis, which took several days. We cannot exclude that larger repositories could exhibit different patterns of micro-changes or refactoring activities. Our results may therefore not fully represent the diversity of software development practices in projects of significantly larger scale, potentially affecting the applicability of our conclusions to such contexts.

\section{Reflection}

The concept of micro-change, as explored in this study, opens up novel avenues for enhancing software development and maintenance practices. Two particularly promising applications of micro-change that we envisage are in the realms of commit message generation and the development of a tool that not only highlights code changes but also annotates them with corresponding micro-change types.

\paragraph*{Commit Message Generation} An automated commit message generation system powered by the concept of micro-changes could significantly improve the efficiency of the version control workflow. By analyzing the micro-changes within a commit, such a system could generate concise, descriptive commit messages that accurately reflect the essence of the code changes. This would not only save developers and reviewers' time for understanding code change but also improve the quality of documentation within repositories, making it easier to understand the history and intent of changes. Such messages could enhance clarity for future maintenance, reviews, and collaborative work by providing a clear and automatically generated summary of the impact of each commit.

\paragraph*{Micro-Change Annotation System} We think that it is possible to build a development tool that goes beyond simply highlighting differences between code versions. Such a tool would leverage the micro-change catalog to annotate changes with specific micro-change types, offering developers immediate insights into the nature of each modification. For instance, a change categorized under \ReverseCondition would be clearly marked as such, providing context at a glance. This could facilitate a more nuanced understanding of code evolution for developers reviewing changes, thus improving code review efficiency and accuracy.

Moreover, by making the implications of changes more transparent, such a tool could aid in educational contexts, helping new developers understand coding practices and patterns through real examples. Both these applications highlight the potential of micro-change to not only advance current development practices but also to foster a deeper understanding of code evolution. As we continue to refine and complete our micro-change catalog and detection methodologies, we look forward to exploring these and other applications of micro-changes.

\section{Conclusion and Future Work}

In this paper, we introduced the concept of micro-changes, transforming code diffs into a series of predefined operations described in natural language. We have developed a catalog tailored to conditional-related micro-changes, complete with specific detection rules. 
Our detection efforts across 73 open-source repositories reveal that, on average, 67.1\% of conditional-related changes are covered by micro-changes.

In the future, we aim to broaden the scope of our catalog beyond conditional-related changes and explore practical applications for leveraging micro-changes.

\section*{Acknowledgments}
This work was partly supported by the Young Researchers Exchange Programme between Japan and Switzerland under the Japanese-Swiss Science and Technology Programme, JST SPRING No.\ JPMJSP2106, and JSPS Grants-in-Aid for Scientific Research Nos.\ JP24H00692, JP23K24823, JP21H04877, JP21K18302, and JP21KK0179.

\IEEEtriggeratref{25}

\end{document}